\documentclass[sn-mathphys,Numbered]{sn-jnl}

\usepackage{graphicx}%
\usepackage{multirow}%
\usepackage{amsmath,amssymb,amsfonts}%
\usepackage{orcidlink}
\usepackage{amsthm}%
\usepackage{mathrsfs}%
\usepackage[title]{appendix}%
\usepackage{xcolor}%
\usepackage{textcomp}%
\usepackage{manyfoot}%
\usepackage{booktabs}%
\usepackage{algorithm}%
\usepackage{algorithmicx}%
\usepackage{algpseudocode}%
\usepackage{listings}%

\begin{document}

\title[Article Title]{Gravitational Waves on Charged Black Hole Backgrounds in Modified Gravity}


\author*[1]{\fnm{Miguel} \sur{Barroso Varela \orcidlink{0009-0006-9844-7661}}}\email{mb3119@imperial.ac.uk}

\author*[1]{\fnm{Hugo} \sur{Rauch \orcidlink{0009-0002-2380-7280}}}\email{har119@imperial.ac.uk}

\affil[1]{\orgdiv{Blackett Laboratory}, \orgname{Imperial College}, \orgaddress{\city{London}, \postcode{SW7 2AZ}, \country{United Kingdom}}}

\abstract{The stability of Reissner-Nördstrom black holes with an extremal mass-charge relation was determined by calculating the propagation speed of gravitational waves on this background in an effective field theory (EFT) of gravity. New results for metric components are shown, along with the corresponding new extremal relation, part of which differs by a global factor of 2 from the past published work. This new relation further develops the existing constraints on EFT parameters. The radial propagation speed for gravitational waves in the Regge-Wheeler gauge was calculated linearly for all perturbations, yielding exact luminality for all dimension-4 operators. The dimension-6 radial speed modifications introduce no constraints on the sign of the modified theory parameters from causality arguments, while the deviation from classical theories vanishes at both horizons. The angular speed was found to be altered for the dimension-4 operators, with possible new constraints on the modified theory being suggested from causality arguments. Results are consistent with existing literature on Schwarzschild black hole backgrounds, with some EFT terms becoming active only in non-vacuum spacetimes such as Reissner-Nördstrom black holes.}




\maketitle

\newpage

\section{Introduction}\label{IntroductionChapter}

This investigation focuses on the stability and physical consistency of charged black holes with a near extremal mass-charge relation. It's important to note that this kind of physical phenomenon is yet to be observed experimentally, which reduces this investigation to a purely conceptual exercise \cite{ExtremalObservation}. Nevertheless, this is a rich environment to study different advanced theories which are not always obviously inconsistent or constrained by their compatibility with classical results. This is why the topic of extremal black holes has been so thoroughly researched in the past and even presently \cite{Extremal,ExtremalBlackHoles}, with applications ranging all the way to connections to dark matter in the primordial universe \cite{ExtremalDarkMatter}. \par
This kind of black hole presents us with an unprotected point of infinite curvature at its centre, which is severely incompatible with our current understanding of physics and therefore pushes us towards investigating any physical instabilities that could point to these black holes being kept from ever existing \cite{RNStability,RNStability2}. Our investigation aims to analyse the effect that an effective field theory of gravity has on this kind of phenomenon, with the propagation of gravitational waves being employed as a probing mechanism \cite{BlackHoleWaves}. Using stability and causality arguments, we intend to determine any pathologies that could provide us with insight into extremal black holes and perhaps even establish constraints on the coefficients of the perturbations in these kinds of modified theories of gravity. \par
Importantly, this is by no means an unfounded attempt, as research employing these types of gravitational wave causality and stability arguments to investigate different aspects of modified theories of gravity has been around for several years. This includes literature where the changes implied by a modified theory on the charged black hole metric were determined \cite{ExtremalBlackHoles} and an investigation into wave perturbations to uncharged black hole backgrounds \cite{BlackHoleWaves}, among many other applications \cite{VacuumCausality,GWProbe}. Modified theories of gravity themselves have been a growing topic over the past decades, with many proposed alternatives to Einstein's original theory, such as f(R) gravity \cite{FRGravity,NMCGravity}, patching profound issues in the current cosmology and quantum gravity landscape \cite{HubbleFR,HubbleModifiedGrav}.  \par
The layout of this paper is as follows. We discuss the effective field theory of gravity on which we focused for our research, with details on the perturbed stress-energy tensor and Maxwell equations given in Section \ref{EFTChapter}. The effects of this modified theory on the black hole metric and its electric field are presented in Section \ref{PerturbedMetricChapter}, along with the consequences these present on the extremal relation. This is followed by a discussion of the methods employed to analyse the propagation of gravitational waves and massless scalar fields on classical and perturbed charged black hole backgrounds, which is included in Section \ref{SpeedChapter}. We finalise this paper in Section \ref{ResultsChapter}, where we present the obtained results, draw conclusions from them and suggest possible extensions to our work. Details on some of the longer results are given in the Appendix. We work with the $(-,+,+,+)$ signature, choose units where $c=\epsilon_0=\mu_0=1$ and define $8\pi G=\kappa^2$.

\section{Effective Field Theory of Gravity} \label{EFTChapter}


\subsection{Perturbed Stress-Energy Tensor}
Although theories of quantum gravity add new heavy fields to our physical theories, leading to extremely non-trivial dynamics, they may be integrated out at low energy scales and represented through their effects on the classical fields \cite{BlackHoleWaves}. With this in mind, throughout this investigation, we analyse a low-energy effective field theory (EFT) of gravity described by a perturbed EM Lagrangian which is altered by all possible dimension-4 operators \cite{ExtremalBlackHoles} and all vacuum dimension-6 operators \cite{BlackHoleWaves}. This is given by 
\begin{equation} \label{PerturbedAction}
\begin{aligned}
\mathcal{L}= & \sqrt{-g} \left( \frac{R}{2 \kappa^2}-\frac{1}{4} F_{\mu \nu} F^{\mu \nu}\right)+\mathcal{L}_{D4}+\mathcal{L}_{D6},
\end{aligned}
\end{equation}
where the higher dimensional operators are given by
\begin{equation}
\begin{aligned}
\mathcal{L}_{D4}=\sqrt{-g} &\left[  c_1 R^2+c_2 R_{\mu \nu} R^{\mu \nu}+c_3 R_{\mu \nu \rho \sigma} R^{\mu \nu \rho \sigma}\right. \\
 & +c_4 R F_{\mu \nu} F^{\mu \nu}+c_5 R^{\mu \nu} F_{\mu \rho} F_\nu{ }^\rho+c_6 R^{\mu \nu \rho \sigma} F_{\mu \nu} F_{\rho \sigma}+c_7\left(F_{\mu \nu} F^{\mu \nu}\right)^2 \\
 & \left.+c_8\left(\nabla_\mu F_{\rho \sigma}\right)\left(\nabla^\mu F^{\rho \sigma}\right)+c_9\left(\nabla_\mu F_{\rho \sigma}\right)\left(\nabla^\rho F^{\mu \sigma}\right)\right]
 \end{aligned}
\end{equation}
and 
\begin{equation}
\begin{aligned}
\mathcal{L}_{D6}=  \sqrt{-g}& {\left[d_1 R \square R+d_2 R_{\mu \nu} \square R^{\mu \nu}+d_3 R^3+d_4 R R_{\mu \nu}^2\right.} \\
& +d_5 R R_{\mu \nu \alpha \beta}^2+d_6 R_{\mu \nu}^3+d_7 R^{\mu \nu} R^{\alpha \beta} R_{\alpha \nu \mu \beta}+d_8 R^{\mu \nu} R_{\mu \alpha \beta \gamma} R_\nu^{\alpha \beta \gamma} \\
& \left.+d_9 R_{\mu \nu}{ }^{\alpha \beta} R_{\alpha \beta}{ }^{\gamma \sigma} R_{\gamma \sigma}{ }^{\mu \nu}+d_{10} R_\mu{ }^\alpha{ }_\nu{}^\beta R_\alpha{}^\gamma{ }_\beta{}^\sigma R_\gamma{ }^\mu{ }_\sigma{ }^\nu\right].\\
&
\end{aligned}
\end{equation}
As we will be investigating extremal black holes near their horizon, we expect $r \sim m\sim q$, where $m=M/4\pi$ and $q=Q/4\pi$ and the charge and mass of the black hole are $Q$ and $M$ respectively. The $c_i$ perturbations are dimension-4 in the sense that they are of order $1/q^4$, which is clear from the $c_7$ term (as each $F$ is proportional to $q/r^2\sim 1/q$). Due to $r\sim q$ and the form of the metric, each derivative adds a factor of $1/q$, which causes any curvature objects (which include double derivatives of the metric) to be $\sim 1/q^2$.  The original terms in the Lagrangian are therefore clearly of lower (second) order, being proportional to $1/q^2$, and thus are less suppressed. The remaining dimension-6 operators would include, for example, other combinations such as $R^2 F^{\mu\nu}F_{\mu\nu}$, but are also highly suppressed. The $c_i$ and $d_i$ coefficients are expected to include suppression due to the scale of the mass of the abstract heavy field that would give rise to these Lagrangian terms \cite{ExtremalBlackHoles,SpeedOfGravity,SubluminalWaves}.  \par

The perturbed field equations are given by
\begin{equation}
G^{\mu \nu}=\frac{2\kappa^2}{\sqrt{-g}}\left(\frac{\delta (\sqrt{-g}\mathscr{L}_{\text {EM}})+\sum_i \delta(\sqrt{-g} \mathscr{L}_{c_i})}{\delta g_{\mu \nu}}\right)=\kappa^2\left( T_{\text{EM}}^{\mu\nu}+ \Delta T^{\mu\nu}\right),
\end{equation}
where the first term on the right side is the classical electromagnetism $T^{\mu\nu}$ and where we have defined the new term in the stress-energy tensor caused by all the EFT operators. The field equations for some of the terms considered here are presented in \cite{ExtremalBlackHoles,BlackHoleWaves}. The remaining modifications are presented in Appendix \ref{FieldEquationsAppendix}.
\par
The linearly altered equations of motion cause a linear change in the Reissner-Nördstrom (RN) metric, which in the notation of the remaining sections of this paper can be written as $g_{\mu\nu}=\bar g_{\mu\nu}+ \sum \delta g_{\mu\nu}$. Importantly, all terms in $\Delta T^{\mu\nu}$ should be calculated using the zeroth order (unperturbed) quantities, as they are already being multiplied by $c_i$ and hence are already of the highest possible order. Unperturbed quantities will always be denoted with a bar ($\bar F^{\mu\nu}$, for example) when the distinction is necessary, but one should always be aware that any quantity already multiplied by $c_i$ should be calculated to zeroth order only \cite{ExtremalBlackHoles}.

\subsection{Perturbed Maxwell Equations}
Similarly to the work described in the previous section, we must also consider the changes the perturbation may induce in the Maxwell equations \cite{ExtremalBlackHoles}. This is especially true for the $c_i$ terms which we analyse in this paper, as some include $F$ in their expression in the Lagrangian. By explicitly perturbing the Euler-Lagrange equations for the EM vector field we arrive at the new Maxwell equations
\begin{equation}
\nabla_\mu F^{\mu \nu}=\bar{\nabla}_\mu\left(\sum_i c_i J_i^{\mu\nu}\right),
\end{equation}
where $J_i^{\mu\nu}$ is some resulting quantity from each of the coefficients in the Lagrangian, as shown in \cite{ExtremalBlackHoles}. We also note the right side of the relation is explicitly linear in $c_i$. Therefore, the same precaution discussed previously must be taken when determining the quantities to lowest order. As expected, when we set $c_i=0$ we obtain the classical equations, which is a good test of the method employed in this derivation. None of the $d_i$ coefficients introduce any changes to this expression, due to the absence of any EM dependence in their Lagrangian term.\par
This new relation indicates that we could also have some changes to our Maxwell tensor, which would be manifested as changes in the electric and magnetic fields of the system. However, as all perturbations to the Lagrangian should be evaluated to zeroth order (as they are already being multiplied by $c_i$), we expect all new quantities to preserve the same spherical symmetry initially obeyed by the unperturbed background. This motivates us to expect simple changes to our EM fields, with only the radial component of the electric field (already present in the unperturbed background) being altered, while all other EM field components remain unaltered (non-existent). By definition, the change in the Maxwell tensor must also obey the antisymmetric properties of its original counterpart. Considering all this, we define the perturbed quantity as $F^{\mu\nu}=\bar F^{\mu\nu}+c_i f_i^{\mu\nu}$, where $f_i^{tr}=-f_i^{rt}\equiv\Delta E_i (r)$. Our new Maxwell equations then become
\begin{equation}
\nabla_\mu F^{\mu \nu}=\nabla_\mu \bar{F}^{\mu \nu}+c_i \bar{\nabla}_\mu f_i^{\mu\nu}=c_i \delta\Gamma^\mu_{\mu\beta}\bar{F}^{\beta\nu}+c_i\bar{\nabla}_\mu f_i^{\mu\nu},
\end{equation}
where we have used the antisymmetry of the Maxwell tensor and the symmetry of the Christoffel symbols. Finally, we have denoted the perturbation to the covariant derivative as the original operator plus an additional alteration to the Christoffel symbol $\delta \Gamma$, which is caused by the linear modifications to the metric.

\section{Perturbed Metric and Extremal Relation} \label{PerturbedMetricChapter}

In this section, we describe the process of determining the perturbations to the time component of the metric and the shift in the electric field, which are central to the result of that paper, while never being explicitly presented in that publication. Most of these calculations, along with many others in the remaining sections of this paper, were performed in the Mathematica software using code developed especially for this work. This code was first checked with known results from classic GR, such as the Schwarzschild metric, to confirm the correct implementation of all differential geometry quantities, along with the field equations themselves. Thus, the results of this section also serve as an additional confirmation of the foundations of our code, as the methods employed here are far from trivial and any issues would become immediately obvious from the resulting equations. 

\subsection{Metric and Electric Field Perturbations} 
\subsubsection{Perturbations for \texorpdfstring{$c_i$}{Lg} Coefficients}
As we assume a static spherically symmetric background, we may use the ansatz 

\begin{equation}
    ds^2=-\left(\bar{A}+c_i \delta A_{c_i}\right)dt^2+\frac{1}{\bar{B}+c_i \delta B_{c_i}}dr^2+r^2 d\theta^2+r^2 \sin^2 \theta d\phi^2,
\end{equation}
where we write the classic components of the RN metric as $\bar A=\bar B=1-\frac{\kappa^2 m}{r}+\frac{\kappa^2 q^2}{2 r^2}$. The spherical symmetry of the system allows us to solve the field equations for a general stress-energy tensor and determine the values of these functions in the metric using the integrals \cite{MetricIntegrals}

\begin{equation} \label{ABIntegrals}
    B=1-\frac{\kappa^2 m}{r}-\frac{\kappa^2}{r} \int_{r}^{\infty} dr \ r^2 T_0^0  
     \quad\quad\quad\quad\quad
      A=B \ \exp \left(\kappa^2\int_r^{\infty} dr \ \frac{r}{B}\left(T_0^0-T_1^1\right)\right).
\end{equation}
For the radial component of the metric this is
\begin{equation} \label{PerturbedBIntegral}
B=1-\frac{\kappa^2 m}{r}+\frac{\kappa^2 q^2}{2 r^2}-\frac{c_i \kappa^2}{r} \int_r^{\infty} dr \ \delta T^0_0=\bar{B}-\frac{c_i \kappa^2}{r} \int_r^{\infty} dr \ \delta T^0_0
\end{equation}
and similarly for the time component
\begin{equation}
\begin{aligned}
  A=& B \ \exp \left(\kappa^2 \int_r^{\infty} dr \ \frac{r}{\bar{B}}\left(1-\frac{c_i \delta B}{\bar{B}}\right)c_i\left(\delta T_0^0-\delta T_1^1\right)\right)= \\
  =& B \ \exp \left(c_i \kappa^2 \int_r^{\infty} dr \ \frac{r}{\bar{B}}\left(\delta T_0^0-\delta T_1^1\right)\right)= \\
  =&B \left(1+c_i \kappa^2 \int_r^{\infty} dr \ \frac{r}{\bar{B}}\left(\delta T_0^0-\delta T_1^1\right)\right)= \\
  =& B+ c_i \kappa^2 \bar{B}\int_r^{\infty}dr \ \frac{r}{\bar{B}}\left(\delta T_0^0-\delta T_1^1\right) + \mathscr{O}(c^2),
\end{aligned}  
\end{equation}
where we have only kept terms which are linear in $c_i$, in line with our perturbative treatment of that coefficient. We have also used the fact that we must be able to write the stress-energy tensor as we did for the RN spacetime plus a linear perturbation, which we define as $\delta T_{\mu\nu}$. The zeroth order part of this tensor obeys $\bar{T}^1_1=\bar{T}^0_0$ and will necessarily integrate to give the original components of the metric. \par
It's important to note that the perturbation to the classical EM stress-energy tensor is not uniquely $c_i\Delta T_{\mu\nu}$. There is an additional contribution from the undetermined perturbation to the Maxwell tensor and from the perturbed metric, which is used to raise and lower indices in order for us to be consistent with the version of the Maxwell tensor we use. We choose this to be $F^{\mu\nu}$ in order to align our approach with the one in \cite{ExtremalBlackHoles}. This yields the full EM stress-energy tensor
\begin{equation}
\begin{aligned}
 (T_{EM})_{\ \nu}^\mu=&F^{\mu\gamma}F_{\nu\gamma}-\frac{1}{4}F^{\alpha\beta}F_{\alpha\beta}\delta^\mu_{\nu}= \\
 =&\left(\bar{F}^{\mu \gamma}+c_i f_i^{\mu \gamma}\right)\left(\bar{F}^{\alpha \beta}+c_i f_i^{\alpha \beta}\right) g_{\nu \alpha} g_{\gamma \beta} \\
 &-\frac{1}{4}\left(\bar{F}^{\rho\sigma}+c_i f_i^{\rho\sigma}\right)\left(\bar{F}^{\alpha \beta}+c_i f_i^{\alpha \beta}\right) g_{\alpha \rho} g_{\beta \sigma} \delta_\nu^\mu,
\end{aligned}
\end{equation}
where the metric $g_{\mu\nu}$ will also include a perturbation which we define below. Thus, we define the full stress-energy tensor as 
\begin{equation}
    T^\mu_{\ \nu}=(\bar{T}_{EM})^\mu_{\ \nu}+c_i\delta T^\mu_{\ \nu},
\end{equation}
where we have now completely separated the zeroth and first-order contributions. 
\par 
Interestingly, we find that the two components of the stress-energy tensor differ by a term proportional to $\bar A$, which implies that they are exactly the same when evaluated at the old horizon radii. This will be an important point when considering the new horizon radii. Evaluating the 4 components of the Maxwell equations, we find that only one is not automatically satisfied. Additionally, it is fully first order, as all zeroth order components necessarily cancel out, allowing us to solve for the electric field shift. Namely, for the $c_6$ coefficient this gives
\begin{equation}
    \Delta E_{c_6}=-\frac{15 \kappa^2 q^3}{r^6}+\frac{8\kappa^2 qm}{r^5},
\end{equation}
which obeys several conditions that we should expect for our electric field shift. Namely, it goes to 0 when we move infinitely away or when the charge goes to 0. These are both sensible, as we should feel no effect of the black hole when infinitely far away and as our Lagrangian perturbation is proportional to the Maxwell tensor, which disappears when we remove all the charge from the system. However, we find that this and all of the remaining $c_i$ electric field shifts have no effect on the speed of gravitational waves on this background and are hence disregarded from this point forward. \par
With all other unknowns found, we plug the resulting function of $r$ into the integrals above and find the perturbations to the metric, which match those in \cite{ExtremalBlackHoles}. These show an interesting effect of the perturbation, as the time and the radial components are no longer the same, matching each other only at the original event horizon radii $R_{\pm}$. 

\subsection{Perturbations for \texorpdfstring{$d_i$}{Lg} Coefficients}
The same methodology can be applied to the dimension-6 perturbations of the action, with the modified field equations for some of these determined in \cite{BlackHoleWaves}. However, that paper considered the spacetime around an uncharged black hole, which leads to the zeroth order equation $\bar R_{\mu\nu}=0$, causing any terms that are at least quadratic in $\bar R_{\mu\nu}$ to vanish. The same is not true for our considerations, as we are no longer in a vacuum. Hence, all of these perturbations are unique to $q\neq 0$, with some only arising for non-vacuum backgrounds. \par 
The absence of any EM dependence on the perturbations considered here means that we don't need to consider changes to the electric field as before. Following the same steps as above, we first obtain the $A-B$ relation for each coefficient, which may then be applied to $\delta T^0_0$. Plugging this into the integrals in Equation \ref{ABIntegrals}, we extract the radial metric component and consequently its time counterpart. The results for all of these are presented in Appendix \ref{MetricModifications}.
All of these results match the ones in \cite{BlackHoleWaves} when $q=0$, due to the simple relation between RN and Schwarzschild black holes. However, while the uncharged results can be obtained simply from the above, the opposite is not possible.
 \subsection{Perturbations to Extremal Relation}\label{ExtremalRelation}
The perturbed metric components could lead to different radii for the event horizons, as we must now impose new conditions $A(r_H)=B(r_H)=0$, which must be satisfied simultaneously, therefore leading to 2 equations to solve for $r_H$. We know that to zeroth order $\bar{A}(r)=\bar{B}(r)$, while $\delta A(r)$ and $\delta B(r)$ clearly aren't the same function in general. However, as both of these perturbations are already multiplied by $c_i$, we may evaluate them both at the classical horizon radii $R_{\pm}$. Careful evaluation shows that these then become exactly the same function, meaning that we only have a single independent equation to solve for our single unknown $r_H$. We therefore compute the unperturbed metric component at the linearly perturbed horizon radius 

\begin{equation}
\begin{aligned}
& \bar{A}(R_{\pm}+c\Delta r_{\pm})=1-\frac{\kappa^2 m}{R_{\pm}+c \Delta r_{\pm}}+\frac{\kappa^2 q^2}{2(R_{\pm}+c \Delta r_{\pm})^2}= \\
& =1-\frac{\kappa^2 m}{R_{\pm}}+\frac{\kappa^2 q^2}{2 R_{\pm}^2}+c \kappa^2 \Delta r_{\pm} \left(\frac{m}{R_{\pm}^2}-\frac{q^2}{R_{\pm}^3}\right)= \\
& =0+\frac{c \kappa^2}{R_{\pm}^3} \Delta r_{\pm}\left(mR_{\pm}-q^2\right),
\end{aligned}
\end{equation}
which, together with the result for $y(r)$ evaluated at the zeroth order horizon radius, allows us to solve for the perturbation to this radius as
\begin{equation}
    r_{\pm}=R_{\pm}+\sum_i c_i \frac{R_{\pm}^3}{\kappa^2\left(q^2-m R_{\pm}\right)}\delta A_i(R_{\pm}).
\end{equation}
By imposing $r_+=r_-$ we obtain the modified $m/q$ relation for extremality. Plugging in all the different coefficients and combining all of these linearly, we construct the new extremal relation
\begin{equation}
\begin{aligned}
\frac{\kappa}{\sqrt{2}} \frac{m}{|q|}=1-\frac{1}{5 q^2} &\left( 2c_2+8 c_3+\frac{ 2c_5}{\kappa^2}+\frac{ 2c_6}{\kappa^2}+\frac{8 c_7}{\kappa^4}-\frac{ 2c_8}{\kappa^2}-\frac{c_9}{\kappa^2}\right. \\
& \left.-\frac{100 d_2}{63 \kappa^2 q^2}-\frac{680 d_5}{63 \kappa^2 q^2}-\frac{10 d_6}{3 \kappa^2 q^2}+\frac{10 d_7}{3 \kappa^2 q^2}\right.\\
&\left.-\frac{100 d_8}{63 \kappa^2 q^2}+\frac{160 d_9}{21 \kappa^2 q^2}+\frac{10 d_{10}}{21 \kappa^2 q^2}\right).
\end{aligned}
\end{equation}
The dimension-4 part precisely matches the expression in \cite{ExtremalBlackHoles}, apart from a total factor of 2 by which our result differs on all $c_i$ perturbations. We should note that the result in the original paper was merely used to impose a condition on the sign of the entire linear combination of coefficients within the parentheses, which is not altered by having an extra general factor of 2. Nevertheless, the similarity between both results is a good check that we have followed the right procedure. Although only $R^3$ operators are considered for the $d_i$ coefficients in our EFT, these new terms provide additional insight into the argument in \cite{ExtremalBlackHoles}. The same constraint can now be written as
\begin{equation}
  \begin{aligned}
   &2c_2+8 c_3+\frac{ 2c_5}{\kappa^2}+\frac{ 2c_6}{\kappa^2}+\frac{8 c_7}{\kappa^4}-\frac{ 2c_8}{\kappa^2}-\frac{c_9}{\kappa^2}-\frac{100 d_2}{63 \kappa^2 q^2}-\frac{680 d_5}{63 \kappa^2 q^2}-\frac{10 d_6}{3 \kappa^2 q^2}\\
   &+\frac{10 d_7}{3 \kappa^2 q^2}-\frac{100 d_8}{63 \kappa^2 q^2}+\frac{160 d_9}{21 \kappa^2 q^2}+\frac{10 d_{10}}{21 \kappa^2 q^2} \geq 0.
   \end{aligned}
\end{equation}
Apart from the dimension-6 modifications to the relation being suppressed by the energy scale of the EFT via the $d_i$ coefficients, the additional factor of $\kappa^2 q^2$ increases their relevance, as this is expected to be a small value for the very lightly charged black holes one would expect from charge neutrality arguments. This extremal relation is the one we should impose when analysing our final results for the gravitational wave speeds, as this is the new condition for the creation of a charged black hole with a naked singularity in the altered theory of gravity.

\section{Scalar Field and Gravitational Wave Speed} \label{SpeedChapter}

The main idea of this investigation is to investigate the propagation speed of gravitational waves on a perturbed RN black hole background. When looking for pathologies in this result, we'll be particularly interested in its comparison with the speed of light $c$ (=1 in the units we have initially imposed) \cite{BlackHoleWaves}. Any slower propagation, labelled as subluminal, is perfectly acceptable from a causality point of view, as it should by no means break the universal limit \cite{SpeedOfGravity}. However, if we find it to have the opposite behaviour (superluminal), then we must conduct further analysis in order to determine if there are any inconsistencies with the expected physical principles of causality, as there still might not be any fundamental issues with our result \cite{SubluminalWaves}. \par
Throughout the remaining sections of this paper, we will be interested in extracting the propagation speed of different fields in our background metric. The first issue is defining what we mean by this speed. In order to analyse causality considerations, we must consider the speed at which information propagates, as that is what any observer would detect. For this, we are interested in the initial bit of information, which is therefore at the front of the wave. Its propagation speed is determined by the front velocity, as described in \cite{MassiveGravity}. Explicitly, this is the large $k$ (or high frequency) limit of the phase velocity, being calculated as $v_f=\lim_{k\rightarrow\infty} \frac{\omega}{k}$. Here $\omega$ is the frequency of the wave and $k$ describes its wavevector. 
Other descriptions of speed, such as phase and group velocity, can and have been observed and measured experimentally to be superluminal, while still causing no contradiction with the pillars of causality \cite{MassiveGravity}. However, if we find a superluminal front velocity, which describes the propagation of new information, we face acausalities and incompatibilities with fundamental physical principles \cite{SpeedOfGravity,SubluminalWaves}. We may therefore use this to uncover issues with the physical systems we are analysing or to possibly impose constraints on the coefficients of our effective field theory of gravity.

\subsection{Determining Speeds from the Effective Metric}

When determining the propagation speed of fields in this investigation we follow the WKB approximation \cite{WKB}, implying that the wavelengths we consider should be short, such that they are no longer than the scale over which our geometry (the black hole) changes. Thus, we may take the highest powers of $\omega$ and $k$ to dominate, as these will be much greater than any lower powers. Nevertheless, in order to obey the energy scale of the low-energy EFT, we should also keep these wavelengths from becoming arbitrarily short, keeping it in between these two regimes, as will become relevant when discussing the resolvability of these effects later on \cite{SubluminalWaves}. By modelling the wave-like behaviour via Fourier expansions in terms of plane waves evolving as $\exp(i\omega t-ikr)$, we may treat time derivatives as factors of $i\omega$ and radial derivatives as factors of $ik$. The previous logic then demands that we investigate the highest orders of these derivatives (and their combinations), with others being negligible for our equations. This so-called characteristic analysis \cite{SpeedOfGravity} will be applied throughout this section, where as long as we can guarantee that all plane wave behaviour (among others we include, such as angular behaviour in spherical harmonics) may be factored out, we can analyse our equations in a much simpler manner. \par
It is useful to define the notion of an effective metric \cite{BlackHoleWaves} via the equation
\begin{equation}\label{EffectiveBox}
    \Box_Z \varphi+U(r)\varphi\equiv Z^{\mu\nu}\nabla_\mu \nabla_\nu \varphi+U(r)\varphi=0,
\end{equation}
where the covariant derivative is defined in terms of $Z_{\mu\nu}$ and the potential is written as $U(r)$. The effective metric $Z_{\mu\nu}$ is defined as 
\begin{equation}  
Z_{\mu\nu}=\text{diag}(-Z_t(r),Z_r^{-1}(r),Z_\Omega(r) r^2,Z_\Omega(r) r^2 \sin^2 \theta),
\end{equation}
with its inverse $Z^{\mu\nu}$ defined as it would for a regular matrix. This defining equation is written in an analogous manner to a wave equation for a massless scalar field, with the flat spacetime $\Box=-\partial^2_t+\nabla^2$ replaced with the $\Box_Z$ operator, which was defined above. Similarly to how a simple wave equation allows us to read off the propagation speed as $v^2\nabla^2\varphi=\partial^2_t \varphi$, this method simplifies the analysis we aim to conduct. Expanding this equation, we obtain
\begin{equation}
\varphi^{\prime \prime}+\frac{1}{2} \varphi^{\prime}\left(\frac{\left(Z_{t} Z_r\right)^{\prime}}{Z_t Z_t}+\frac{2 Z_{\Omega}^{\prime}}{Z_{\Omega}}+\frac{4}{r}\right)-\frac{\ddot{\varphi}}{Z_t Z_r}+\frac{1}{Z_{\Omega} Z_{r} r^2} \nabla_{\Omega}^2 \varphi+\frac{U(r)}{Z_r}\varphi=0,
\end{equation}
where we have denoted time derivatives as $\dot\varphi$, radial derivatives as $\varphi'$ and the angular Laplacian operator as $\nabla^2_{\Omega}$. We may simplify this by assuming a simple spherically symmetric wave form for our arbitrary scalar field \cite{BlackHoleWaves}. We use the ansatz $\varphi(t,r,\theta)=\frac{\phi(r)}{r}e^{-i\omega t}Y_l(\theta)$, where we have exploited the spherical symmetry of the system with the use of the spherical harmonics $Y_l(\theta)$ and ignored the remaining angular dependence, as it may be removed with a simple rotation. These harmonics have a convenient defining equation given by
\begin{equation} \label{SphericalHarmonicEquation}
    \nabla^2_{\Omega}Y_l(\theta)=-l(l+1)Y_l(\theta)=-JY_l(\theta),
\end{equation}
which simplifies the angular Laplacian term acting on $\varphi$. We will also ignore any potential terms in our investigation, as these should be negligible when determining kinematic properties in the high-frequency limit \cite{BlackHoleWaves,SpeedOfGravity}. This simplifies the equation to give
\begin{equation}
    \phi''+\phi'\left(\frac{(Z_t Z_r)'}{2Z_t Z_r}+\frac{Z_\Omega '}{Z_\Omega}\right)+\left(\frac{\omega^2}{Z_t Z_r}-\frac{J}{Z_\Omega Z_r r^2}-\frac{(Z_t Z_r)'}{2r Z_t Z_r}-\frac{Z_\Omega '}{r Z_\Omega}\right)\phi=0,
\end{equation}
allowing us to extract the effective metric components for any given second-order differential equation. \par
To determine the speed from the effective metric, we may use a similar technique to the one applied in classical GR, using an ansatz $e^{ik_\rho x^\rho}$ and hence writing the highest order derivative terms in (\ref{EffectiveBox}) as 
\begin{equation}
Z^{\mu\nu}\partial_\mu\partial_\nu e^{ik_\rho x^\rho}=\left(Z^{\mu\nu}k_\mu k_\nu\right)e^{ik_\rho x^\rho}=\left(Z^{\mu\nu}g_{\nu\sigma}k_\mu k^\sigma \right)e^{ik_\rho x^\rho}=0.
\end{equation}
By analysing both $\mu=0$ and $\mu=i$, with $i$ labelling spatial components, we obtain the dispersion relation 
\begin{equation}
    Z^{t\nu}g_{\nu\sigma}k_t k^\sigma+ Z^{i\nu}g_{\nu\sigma}k_i k^\sigma=Z^{tt}g_{tt}k_t k^t+Z^{ii}g_{ii}k_i k^i=-Z^{tt}g_{tt}\omega^2+Z^{ii}g_{ii}k^2=0,
\end{equation}
where we have assumed both metrics are diagonal, as is the case for all the cases studied in this investigation. The speed in the $x^i$ direction is thus 
\begin{equation}
    c_i^2=\frac{\omega^2}{k_i^2}=\frac{Z^{ii}g_{ii}}{Z^{tt}g_{tt}},
\end{equation}
where there is no implied sum over $i$.
As will become clear in the next section, this ensures that free massless scalar fields with $Z_{\mu\nu}=g_{\mu\nu}$ propagate luminally, thus setting a good comparison for causality discussions for the propagation of gravitational waves. Inserting the usual notation for the components of the EFT-perturbed RN background metric we obtain the radial speed 
\begin{equation}
    c_r^2=\frac{Z^{rr}g_{rr}}{Z^{tt}g_{tt}}=\frac{Z_t Z_r}{AB}
\end{equation}
and 
using a similar logic, this time with angular propagation, we may also obtain the angular speed, given by
\begin{equation}
    c_\Omega^2=\frac{Z^{\theta\theta}g_{\theta\theta}}{Z^{tt}g_{tt}}=\frac{Z_t}{AZ_\Omega}=\frac{Bc_r^2}{Z_rZ_\Omega},
\end{equation}
where we used the previous expression for the radial speed in order to be able to directly extract this result from our equations of motion from the coefficient on the $J$ term. Hence, any second-order equation of motion can be analysed under this interpretation and speeds may be directly extracted.





\subsection{Free Massless Scalar Fields} \label{ScalarFieldSection}
In the remaining parts of this section, we'll be interested in calculating the speed of gravitational waves when propagating on the perturbed RN background. In order to accurately compare their propagation speed to that of light, it will be useful to first discuss a free massless scalar field moving on the same background metric, which we expect to propagate luminally \cite{BlackHoleWaves}. The Lagrangian for a free massless scalar field is 
\begin{equation} 
    \mathcal{L}=\frac{1}{2}\sqrt{-g}g^{\mu\nu}\partial_\mu \phi \partial_\nu \phi,
\end{equation}
which leads to the equation of motion $\Box \phi=g^{\mu\nu}\nabla_\mu\nabla_\nu\phi=0$. By analysing this equation using the methods introduced in the previous section, we immediately recognise $Z_{\mu\nu}=g_{\mu\nu}$, leading us to the conclusion that this field would propagate luminally, as expected. This will serve as a gauge for what we define as luminal propagation, creating a fixed causality scale to which we can compare our gravitational wave speed \cite{SubluminalWaves}. 

\subsection{Gravitational Waves in Perturbed RN Background}
The propagation of gravitational waves over non-vacuum backgrounds is more complex than the simpler Schwarzschild case. When dealing with these non-trivial cases, we often must exploit symmetries of the system using convenient gauge transformations to simplify our analysis. In this section, we investigate wave-like perturbations over our background, which we shall treat linearly as we have done before. We consider a metric of the form $g_{\mu\nu}=g_{\mu\nu}+\varepsilon h_{\mu\nu}$, where $g$ represents the background and the $\varepsilon$ parameter is kept to first order in all calculations. By calculating all the necessary quantities with the added perturbation, including differential objects such as $R$, we may construct the perturbed Einstein tensor $G_{\mu\nu}$. In the Reissner-Nördstrom case, where $T_{\mu\nu}\neq0$, we must take care to consider the same version of $F_{\mu\nu}$ by raising and lowering quantities with the perturbed metric, as we have done in Section \ref{EFTChapter}. By calculating all components of the field equations $G_{\mu\nu}=\kappa^2 T_{\mu\nu}$, we find that they are automatically satisfied to zeroth order in $\varepsilon$, which is expected and once again confirms the accuracy of our code. We are therefore left with a set of differential equations which are fully $\mathscr{O}(\varepsilon)$ and hence we may solve for $h_{\mu\nu}$. \par 
The spherically symmetric background motivates the imposition of spherically symmetric perturbations, as this symmetry shouldn't be broken by these waves. As discussed in \cite{BlackHoleWaves}, a convenient gauge for this type of system is the Regge-Wheeler gauge \cite{ReggeWheeler}, which separates the odd and even parity components of the waves and reduces the number of apparent degrees of freedom, thus aiding in the decoupling of our equations. In this investigation, we investigate the odd parity perturbations, which are relatively simpler and in general are expected to yield the same result as their even counterparts. We therefore consider perturbations of the form 
\begin{equation}
    h_{\mu\nu}=\begin{pmatrix}
0 & 0 & 0 &h_0(r)\\
0 & 0 & 0 & h_1(r)\\
 0   & 0&0 &0        \\
    h_0(r) & h_1(r)& 0&   0   
\end{pmatrix} e^{-i\omega t} \sin{\theta} Y'_l(\theta),
\end{equation}
where $Y'_l(\theta)$ denotes the derivative of the spherical harmonics and hence the angular part guarantees the odd parity \cite{ReggeWheeler}.
\par
When considering waves travelling on the already perturbed background caused by the modified Lagrangian which we introduced in Section \ref{EFTChapter}, we must consider 2 separate types of perturbations to the classical RN metric. One of these are the wave perturbations and the other is made up of all the contributions from the different $c_i$ coefficients. We write this as 
\begin{equation}
    g_{\mu\nu}=\Bar{g}_{\mu\nu}+\sum_i c_i (\delta g_{\mu\nu})_i +\epsilon h_{\mu\nu},
\end{equation}
where all coefficients $c_i$ and $\varepsilon$ are treated to first order. Because of this, each $c_i$ perturbation may be considered independently, which will greatly decrease the computational and mathematical complexity of our investigation. Nevertheless, $c_i \varepsilon$ terms are kept in all calculations, as they are linear in both coefficients and will capture the effects of the perturbed Lagrangian on the gravitational wave propagation. Due to the unbroken spherical symmetry of the system, we still apply the Regge-Wheeler gauge \cite{ReggeWheeler} and continue to analyse the odd parity perturbations. Thus, we may reuse much of the methodology applied in the classical case, with slight modifications due to the effects of each $c_i$. \par
Throughout these computations, we keep the $c_i$ parts of the metric and the electric field shift abstract as $\delta A$, $\delta B$ and $\Delta E$ in the same way as when we were initially solving for them. Additionally, we take care to define the new doubly perturbed raised Maxwell tensor $F^{\mu\nu}$ from the lowered $c_i$ perturbed version, which we raise with the new doubly perturbed metric. This ensures that we start from the fundamental metric-independent form of the EM tensor before raising it with the new metric which has been affected by gravitational waves. Symbolically this is 
\begin{equation}
    F^{\mu\nu}=(g_{c_i+\epsilon})^{\mu\gamma}(g_{c_i+\epsilon})^{\nu\lambda}(g_{c_i})_{\alpha\gamma}(g_{c_i})_{\beta\lambda}(F_{c_i})^{\alpha\beta},
\end{equation}
where we have denoted doubly perturbed metrics with $c_i+\epsilon$ subscripts.\par 
The same procedure can be followed for all different $c_i$ and $d_i$ perturbations, with some introducing new complexities into our calculations. For example, terms in the $d_2$ stress-energy tensor \cite{ExtremalBlackHoles}, such as $\Box^2 R^{\mu\nu}$, will generate derivatives of the metric of up to sixth-order, along with changes to the equations we use to relate $h_1$ and $h_0$ in the classical case, which we are no longer able to directly solve \cite{BlackHoleWaves}. This equation contains sixth-order terms in $k$ and $\omega$, but we can reduce this by recognising that the zeroth order equation of motion was of the form $h''=f(h,h')$. As these higher-order terms are already $\mathscr{O}(c)$, we may use the unperturbed equations to simplify them \cite{SpeedOfGravity}, given that any other terms would be of $\mathscr{O}(c^2)$. 
The zeroth order equations of motion have already been determined when analysing gravitational waves propagating on a classical RN background, similarly to what was done in \cite{BlackHoleWaves}. Thus, any term that wasn't present in this analysis must necessarily be $\mathscr{O}(c)$ and thus may be simplified using the unperturbed equations \cite{SpeedOfGravity}. Therefore, we are able to lower the order of derivatives on terms such as $h^{(3)}$ by writing $h^{(3)}=(h'')'$ and then plugging in our relation from the classical analysis, where we now include potential terms, as they will now have derivatives acting on them. Fourth derivative terms are obtained from differentiating the previous expression for $h^{(3)}$ and substituting any arising third derivatives using the same expression, with analogous procedures being used for higher-order derivatives. We may then extract the relevant speeds from the resulting second-order differential equations. 

\section{Discussion of Results and Conclusion} \label{ResultsChapter}

\subsection{Gravitational Wave Speed Results}
\subsubsection{Radial Speeds}
After careful calculation of the equations of motion of the gravitational wave perturbations for all the EFT coefficients, we are able to extract the effective metric components and determine the new radial and angular speeds of the waves \cite{BlackHoleWaves}. Interestingly, all of the $c_i$ radial speeds come out to be exactly luminal, meaning that the dimension-4 operators introduce no effects on the radial propagation of these perturbations. This result is by no means trivial, as both the radial and time derivative coefficients seem to be completely unrelated at first, while still cancelling out exactly when combined and analysed to first order in the effective metric components, with no need to insert the explicit expression for $\delta A$ or $\delta B$ in any case. This differs from the Schwarzschild case, where the dimension-4 operators can immediately be ruled out with geometrical arguments in 4D spacetime. This result is also independent of the mass-charge relation and of position, meaning we are unable to draw the expected conclusions about the stability of extremal black holes, which don't seem to be forbidden by causality arguments applied to gravitational wave propagation. \par
Furthermore, this imposes no constraints on the $c_i$ coefficients from causality arguments. In the literature, not only have constraints been determined from weak gravity considerations \cite{ExtremalBlackHoles,WeakestForce}, but also from causality itself for a vacuum \cite{VacuumCausality}. However, it should be noted that the latter example was calculated for a different background than the one analysed here, and therefore the Lagrangian perturbation coefficients aren't necessarily equal to their equivalents in a Reissner-Nördstrom black hole background. The lack of effect of these perturbations on the wave speed is already an interesting result, but it is even more non-trivial than expected, as the results for all coefficients were completely independent of the form of the metric components $X$ and $Y$, which were left abstract at all times. This could be caused by the even parity of the electric field, which could perhaps lead to no coupling on the odd parity waves, while still affecting the even parity modes in profound ways. \par
\begin{figure}[h!] 
    \centering
    \includegraphics[width=11.9cm]{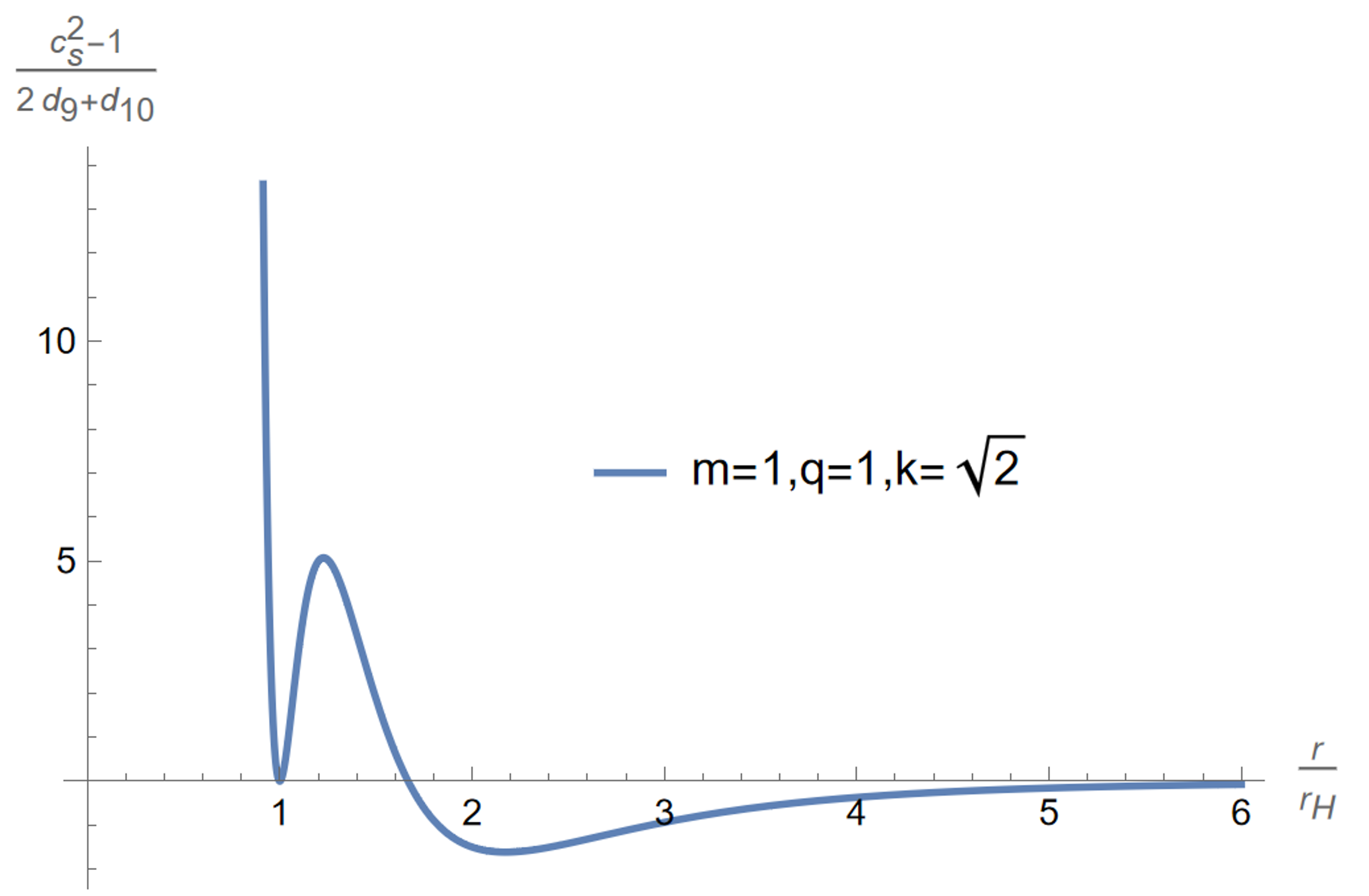}
    \caption{Deviation from luminality of gravitational wave speed near the horizon radius $r_H$, which is unique under the extremal mass-charge relation. Units have been chosen to simplify the interpretation of effects and we have taken $2d_9+d_{10}>0$.}
    \label{SpeedPlot}
\end{figure}

Similarly, most of the $d_i$ terms introduce no speed modifications. However, the coefficients $d_9$ and $d_{10}$ introduce perturbations to the radial speed, as expected from the result on a Schwarzschild background \cite{BlackHoleWaves}. The new speed is
\begin{equation}
\begin{aligned}
c_r^2=1&+\frac{24 d_9 \kappa^2 \bar A\left(-2 \bar A^{\prime}-r^2 \bar A^{(3)}+2 r\bar A^{\prime \prime})\right)}{r^3}\\
&+\frac{3 d_{10}\kappa^2 \bar A \left(12 \bar A-12-8r\bar A^{\prime}+2 r^2 \bar A^{\prime \prime}+r^4 \bar A^{(4)}  \right)}{r^4},
\end{aligned}
\end{equation}
or explicitly
\begin{equation}
c_r^2=1+48 (2d_9+d_{10})\frac{ \kappa^4\left(5 q^2-3 m r\right)\left(r-R_+\right)\left(r-R_-\right)}{r^8},
\end{equation}
which matches the Schwarzschild result when setting $q=0$, as expected. Additionally, this value is exactly luminal at both of the new horizons, as seen from the proportionality to $(r-R_{\pm})$, extending the result from \cite{BlackHoleWaves}, while also being at a min/max (depending on the sign of $2d_9+d_{10}$) point at the outer horizon. The change from subluminal to superluminal (or vice-versa) outside the outer horizon, as seen in Figure \ref{SpeedPlot}, implies that there is no constraint on the sign of $2d_9+d_{10}$ that could guarantee subluminality at all exterior points, unlike in the Schwarzschild case. The precise luminality at the horizon also draws a close parallel with the work in \cite{BlackHoleWaves}, where it is argued that this should be expected from the so-called ``Horizon Theorem". Its validity for the non-vacuum spacetime of RN black holes provides further support to the arguments presented in that publication. 
\par

Given this result, the speed of radial propagation of waves in this and other backgrounds should still be investigated further in future research, as investigating even modes and/or the additional dimension-6 terms could lead to more general conclusions about these kinds of modified theories of gravity. \par

\subsubsection{Angular Speeds}
As no radial speed alterations were found for the dimension-4 terms, we choose to investigate the angular speed of the gravitational waves, for which we obtain 
\begin{equation}
    c_\Omega^2=1+\frac{2\kappa^2 q^2}{r^4}(c_5+2c_6),
\end{equation}
which not only deviates from luminality but also displays interesting features. Globally, this deviation follows a $1/r^4$ dependence, meaning that it quickly decays as we move infinitely away from the black hole. Additionally, it is proportional to $q^2$, which indicates that it is completely independent of the sign of the charge of the black hole, while also being non-existent for a Schwarzschild black hole ($q=0$). Therefore, there is something about the presence of the charge and its consequent radial electric field that causes an effect on the angular speed of waves around the spherically symmetric black hole. Although these perturbations are expected to be very small due to the assumed magnitude of the $c_i$ coefficients in our EFT, we could still impose them to be negative to ensure all waves propagate subluminally. This causality argument could then imply an additional constraint  
\begin{equation}
    c_5+2c_6\leq0,
\end{equation}
which, when combined with the one from \cite{ExtremalBlackHoles}, would provide new insights into acceptable effective field theories of gravity. The physical meaning of this superluminality could be further studied by determining its path difference to luminal motion after a fixed time interval \cite{SubluminalWaves}. However, as we are dealing with a low-energy EFT, we should assume long wavelengths for our particles relative to the energy scale of the theory, thus being much longer than these path differences, which are $\mathscr{O}(c_i)$. These coefficients are highly suppressive in the low-energy EFT and thus the path difference would be far smaller than the wavelength of low-energy particles, making it effectively unresolvable \cite{SubluminalWaves}. Therefore, we don't take these results to point to pathological breaches of causality and such constraints should only be interpreted with this in mind. The same can be argued for the $d_9$ result, which should be taken with the same considerations. \par

\subsection{Conclusion}
This investigation aimed to determine the stability of charged black holes with an extremal mass-charge relation, which exposes their unprotected central singularities to observers at infinity, in a low-energy effective field theory of gravity. The main objective was to use causality arguments applied to the propagation speed of gravitational waves, which would then be analysed at the extremal limit at a radius right outside the horizon to possibly uncover evidence that the non-existence of these physical objects follows directly from quantum gravity, modelled here via all non-zero dimension-4 operators and $R^3$-type dimension-6 operators, hypothesised as the effect on classic fields from undetermined heavy fields added to the classical General Relativity action. In order to achieve this, we first calculated the new perturbed metric, horizon radii and the consequent extremal relation, which involved determining the new stress-energy tensor and Maxwell equations. For those that could be compared, we found the same results as in the literature. Interestingly, our extremal relation perturbation differed from the cited result by a global factor of 2, which wouldn't affect the main conclusion of the original paper but is nevertheless an unexpected deviation. Using the same method, we obtained the new metric for the dimension-6 perturbations, which served as an extension of the same result for Schwarzschild backgrounds conducted in \cite{BlackHoleWaves}. This is especially interesting for all terms which were automatically nullified by the vacuum considered in that work, which are present for our work. \par
The radial speeds for gravitational waves in the new perturbed background all came out to be precisely luminal for the dimension-4 terms in the Lagrangian, while the dimension-6 terms led to deviations from luminality. All of these results were completely independent of the form of the metric perturbations, which were kept abstract throughout the calculation of the equations of motion for the waves, depending only on the relation between the radial and time components of the metric. Not only do these results not lead to any instabilities of the extremal black holes, but they also impose no constraints on the sign of the perturbation coefficients. Comparatively, weak gravity arguments have already been used to determine conditions for some of the same coefficients \cite{ExtremalBlackHoles}, with causality investigations on vacuum backgrounds having found similar relations for analogous coefficients \cite{VacuumCausality}. This work therefore inserts additional terms into this existing constraint. Additionally, the determination of the angular speeds led to other interesting results, with deviations from luminality obtained for 2 of the 9 dimension-4 perturbations. These alterations were found to only exist for charged black holes and could serve as evidence for additional constraints on complete theories of gravity, which could be corroborated by researching the same theory on different backgrounds. Importantly, such conclusions should only be taken into consideration along with the resolvability of such superluminal behaviours in low-energy EFTs \cite{SubluminalWaves}. \par
There are many possible extensions for this investigation. These include:
\begin{itemize}
    \item Repeating the investigation using the even parity perturbations in the Regge-Wheeler gauge. Although these lead to more complex calculations, the even parity of the electric field could lead to there only being significant coupling to even parity waves, which could lead to more profound results.
    \item Analysing an EFT with all remaining dimension-6 operators for Reissner-Nördstrom, including whose effects are not present in Schwarzschild backgrounds \cite{BlackHoleWaves}.
    \item Following the same procedure described in this paper for Kerr (rotating and uncharged) or Kerr-Newmann \cite{KerrNewmanCorrections} (rotating and charged) black hole backgrounds. These may be obtained from Schwarzschild and RN respectively via specific coordinate transformations \cite{KerrTransformation,KerrNewmanTransformation}, which could heavily simplify the calculations for this background by analogy with the work already conducted in published research \cite{ExtremalBlackHoles,BlackHoleWaves} and in this investigation.
    \item Analysing $\mathscr{O}(c^2)$ contributions to the speed and looking for inconsistencies as we approach the new extremal limit. Although these would be highly suppressed by the scale of the EFT, it could be that the linear treatment employed in this investigation removed non-trivial complex behaviour which would be related to the instability of extremal black holes.
    \item If any valid complex behaviour is found in the speed, investigating possible quantum protection mechanisms through which black holes could avoid becoming extremal. These may include, for example, the emission of charged Hawking radiation \cite{NakedEvaporation,HawkingRadiation}.
\end{itemize}

\bmhead*{Acknowledgments}
We'd like to thank Prof. Claudia de Rham for suggesting the analysis of the $c_6$ term, which motivated the remainder of this entire investigation.

\appendix
\section{Metric Perturbations}\label{MetricModifications}
The non-zero metric perturbations for the $d_i$ coefficients are shown below. 

\begin{equation}
\delta A_{d_2} =\frac{118 \kappa^8 {q}^6}{9 {r}^{10}}-\frac{53 \kappa^8 {m} {q^{4 }}}{2 {r}^9}+\frac{4 \kappa^8 {m}^2 {q}^2}{r^8}+\frac{48 \kappa^6 {q}^4}{7 {r}^8}+\frac{24 \kappa^6 {m} {q}^2}{r^7}-\frac{24 \kappa^4 {q}^2}{{r}^6}
\end{equation}

\begin{equation}
\delta A_{d_4} =
\frac{3\kappa^8 m q^4}{r^9}-\frac{2\kappa^8 q^6}{r^10}-\frac{2 \kappa^6 q^4}{r^8}
\end{equation}

\begin{equation}
\begin{aligned}
\delta A_{d_5}=&-\frac{238 \kappa^8 {q}^6}{9 {r}^{10}}+\frac{87 \kappa^8 {m} {q^{4 }}}{{r}^9}-\frac{660 \kappa^8 {m}^2 {q}^2}{7 {r}^8}-\frac{28 \kappa^6 {q}^4}{{r}^8}+\frac{36 \kappa^8 {m}^3}{{r}^7}+\frac{48 \kappa^6 {m} {q}^2}{{r}^7}\\
&-\frac{24 \kappa^6 {m}^2}{r^6}
\end{aligned}
\end{equation}

\begin{equation}
\delta A_{d_{6}} =
\frac{57 \kappa^8 m q^4}{2 r^9}-\frac{179 \kappa^8 q^6}{12 r^{10}}-\frac{27 \kappa^6 q^4}{r^8}
\end{equation}

\begin{equation}
\delta A_{d_{7}} =\frac{20 \kappa^6 q^4}{r^8}+\frac{10 \kappa^8 m^2 q^2}{r^8}+\frac{137 \kappa^8 q^6}{12 r^{10}}-\frac{55 \kappa^8 m q^4}{2 r^9}-\frac{8 \kappa^6 m q^2}{r^7}
\end{equation}

\begin{equation}
\delta A_{d_{8}}=-\frac{115 \kappa^8 {q}^6}{18 {r}^{10}}+\frac{39 \kappa^8 {m} {q}^4}{2 {r}^9}-\frac{151 \kappa^8 {m}^2 {q}^2}{7 {r}^8}-\frac{3 \kappa^6 {q}^4}{{r}^8}+\frac{9 \kappa^8 {m}^3}{{r}^7}+\frac{8 \kappa^6 {m} {q}^2}{{r}^7}-\frac{6 \kappa^6 {m}^2}{{r}^6}
\end{equation}

\begin{equation}
\delta A_{d_{9}} =\frac{93 \kappa^8 m q^4}{7 r^9}-\frac{28 \kappa^8 q^6}{3 r^{10}}-\frac{108 \kappa^8 m^2 q^2}{7 r^8}+\frac{30 \kappa^6 q^4}{r^8}+\frac{10 \kappa^8 m^3}{r^7}-\frac{192 \kappa^6 m q^2}{7 r^7}
\end{equation}

\begin{equation}
\begin{aligned}
\delta A_{d_{10}}=&\frac{53 \kappa^8 q^6}{12 r^{10}}-\frac{507 \kappa^8 m q^4}{28 r^9}+\frac{156 \kappa^8 m^2 q^2}{7 r^8}+\frac{21 \kappa^6 q^4}{2 r^8}-\frac{17 \kappa^8 m^3}{2 r^7}-\frac{138 \kappa^6 m q^2}{7 r^7}\\
&+\frac{9 \kappa^6 m^2}{r^6}
\end{aligned}
\end{equation}

\begin{equation}
   \delta B_{d_{2}}=-\frac{242 \kappa^8 {q}^6}{9 {r}^{10}}+\frac{203 \kappa^8 {m} {q}^4}{2 r^9}-\frac{92 \kappa^8 {m}^2 {q}^2}{r^8}-\frac{680 \kappa^6 {q}^4}{7 {r}^8}+\frac{168 \kappa^6 {m} {q}^2}{r^7}-\frac{72 \kappa^4 q^2}{r^6}
\end{equation}

\begin{equation}
   \delta B_{d_{4}}=\frac{7 \kappa^8 {q}^6}{r^{10}}-\frac{15 \kappa^8 {m} {q}^4}{r^9}+\frac{16 \kappa^6 {q}^4}{r^8}
\end{equation}

\begin{equation}
\begin{aligned}
   \delta B_{d_{5}}=&\frac{896 \kappa^8 {q}^6}{9 {r}^{10}}-\frac{357 \kappa^8 {m} {q^{4 }}}{{r}^9}+\frac{2616 \kappa^8 {m}^2 {q}^2}{7 {r}^8}+\frac{224 \kappa^6 {q}^4}{{r}^8}-\frac{132 \kappa^8 {m}^3}{{r}^7}\\
   &-\frac{336 \kappa^6 {m} {q}^2}{{r}^7}+\frac{144 \kappa^6 {m}^2}{r^6}
\end{aligned}
\end{equation}

\begin{equation}
   \delta B_{d_{6}}=\frac{3\kappa^8 m q^4}{2 r^9}-\frac{17 \kappa^8 q^6}{12 r^{10}}
\end{equation}

\begin{equation}
   \delta B_{d_{7}}=-\frac{43 \kappa^8 {q}^6}{12 {r}^{10}}+\frac{21 \kappa^8 m {q}^4}{2 {r}^9}-\frac{6 \kappa^8 {m}^2 {q}^2}{{r}^8}-\frac{10 \kappa^6 {q}^4}{{r}^8}+\frac{8 \kappa^6 m {q}^2}{{r}^7}
\end{equation}

\begin{equation}
\begin{aligned}
   \delta B_{d_{8}}=&\frac{316 \kappa^8 q^6}{9 r^{10}}-\frac{239 \kappa^8 m q^4}{2 r^9}+\frac{780 \kappa^8 m^2 q^2}{7 r^8}+\frac{80 \kappa^6 q^4}{r^8}-\frac{33 \kappa^8 m^3}{r^7}-\frac{104 \kappa^6 m q^2}{r^7}\\
   &+\frac{36 \kappa^6 m^2}{r^6}
\end{aligned}
\end{equation}

\begin{equation}
\begin{aligned}
   \delta B_{d_{9}}=&\frac{431 \kappa^8 {q}^6}{3 {r}^{10}}-\frac{471 \kappa^8 {m} {q^{4 }}}{{r}^9}+\frac{2766 \kappa^8 {m}^2 {q}^2}{7 {r}^8}+\frac{336 \kappa^6 {q}^4}{{r}^8}-\frac{98 \kappa^8 {m}^3}{{r}^7}\\
   &-\frac{384 \kappa^6 {m} {q}^2}{{r}^7}+\frac{108 \kappa^6 {m}^2}{{r}^6}
\end{aligned}
\end{equation}

\begin{equation}
   \delta B_{d_{10}}=\frac{13 \kappa^8 q^6}{6 r^{10}}-\frac{27 \kappa^8 m q^4}{4 r^9}+\frac{57 \kappa^8 m^2 q^2}{14 r^8}+\frac{6 \kappa^6 q^4}{r^8}+\frac{\kappa^8 m^3}{2 r^7}-\frac{6 \kappa^6 m q^2}{r^7}
\end{equation}

\section{Field Equations}\label{FieldEquationsAppendix}
The modified stress-energy tensor terms not shown in present literature and derived for this work are given below. Note that any term in these equations that is proportional to the Ricci Scalar $R$ has been ignored as to zeroth order in RN backgrounds $\bar R=0$. This effectively nullifies any modification caused by $d_1$ and $d_3$.
\begin{equation}
\begin{aligned}
   \Delta T^{\mu\nu}=&d_2 \left[ R_{\alpha \beta} \square R^{\alpha \beta}-2 R^\mu{}_\gamma\square R^{\nu \gamma} -2 R^\nu{}_\gamma\square R^{\mu \gamma}  \right. \\
   & \left. +2 \nabla_\gamma \nabla^\mu \square R^{\nu \gamma}+2\nabla_\gamma \nabla^\nu \square R^{\mu \gamma}+2g^{\mu \nu} \nabla_\gamma \nabla_\alpha \square R^{\gamma \alpha}-2\square^2 R^{\mu \nu}\right] \\
   &+d_4 \left[-2 R_{\alpha\beta}^2 R^{\mu \nu}-2 g^{\mu \nu} \square R_{\alpha\beta}^2 +2 \nabla^\mu \nabla^\nu R_{\alpha\beta}^2\right]\\
   &+d_6 \left[g^{\mu \nu} R_{\alpha \beta}^3  -6 R^{\nu \gamma} R^{\mu \beta} R_{\beta \gamma}+3 \nabla_\gamma \nabla^\mu\left( R^{\beta \nu} R_\beta^\gamma\right)+3 \nabla_\gamma \nabla^\nu \left( R^{\beta \mu} R_\beta^\gamma\right) \right.\\ 
   & \left. +3 g^{\mu \nu}\nabla_\gamma \nabla_\alpha \left( R^{\beta \gamma} R_\beta^\alpha \right)-3 \Box \left(R^{\beta \mu}R_\beta^\nu\right) \right] \\
   & +d_7\left[g^{\mu \nu} R^{\sigma \gamma} R^{\alpha \beta} R_{\alpha \gamma \sigma \beta}-4 R^{\sigma \gamma} R^\lambda{}_{\gamma \sigma}{}^\mu R_\lambda^\nu-4 R^{\sigma \gamma} R^\lambda{}_{\gamma \sigma}{}^\nu R_\lambda^\mu \right.\\
& +2 \nabla_\gamma \nabla_\alpha\left(g^{\alpha \mu} R^{\beta\sigma } R^\gamma{}_{\beta \sigma}{}^\nu +g^{\alpha \nu} R^{ \beta\sigma} R^\gamma{}_{\beta \sigma}{}^\mu+g^{\mu \nu} R^{\beta \sigma} R^\gamma{ }_{\beta \sigma}{}^\alpha\right. \\
& \left.-g^{\alpha \gamma} R^{\sigma \beta} R^\mu{}_{\beta \sigma}{}^\nu\right)+R^{\sigma \alpha} R^{\nu \beta} R^\mu{}_{\alpha \sigma \beta}+R^{\sigma \alpha} R^{\mu \beta} R^\nu{}_{\alpha \sigma \beta} \\
& \left.+2 \nabla_\gamma \nabla_\beta\left(R^{\mu \nu} R^{\beta \gamma}\right)-\nabla_\gamma \nabla_\beta\left(R^{\gamma\mu} R^{\beta \nu}\right) -\nabla_\gamma \nabla_\beta\left(R^{\gamma\nu } R^{\beta \mu}\right)\right]
\end{aligned}
\end{equation}


\bibliography{References}

\end{document}